\documentclass[prd,twocolumn]{revtex4} 
\usepackage{amsmath}

\begin{document}
\def\prg#1{\medskip\noindent{\bf #1}}       \def\ra{\rightarrow}
\def\lra{\leftrightarrow}                   \def\Ra{\Rightarrow}
\def\nin{\noindent}                         \def\pd{\partial}
\def\dis{\displaystyle}                     \def\inn{\,\rfloor\,}
\def\Lra{{\Leftrightarrow}}                 \def\ads{(A)dS}
\def\Leff{\hbox{$\mit\L_{\hspace{.6pt}\rm eff}\,$}}
\def\nb{\marginpar{\huge ?}}                \def\grl{{GR$_\Lambda$}}
\def\mb{{\rm MB}}                           \def\cs{{\rm CS}}

\def\G{\Gamma}        \def\S{\Sigma}        \def\L{{\mit\Lambda}}
\def\a{\alpha}        \def\b{\beta}         \def\g{\gamma}
\def\d{\delta}        \def\m{\mu}           \def\n{\nu}
\def\th{\theta}       \def\k{\kappa}        \def\l{\lambda}
\def\vphi{\varphi}    \def\ve{\varepsilon}  \def\p{\pi}
\def\r{\rho}          \def\Om{\Omega}       \def\om{\omega}
\def\s{\sigma}        \def\t{\tau}          \def\eps{\epsilon}
\def\nab{\nabla}      \def\btz{{\rm BTZ}}   \def\heps{\hat\eps}
\def\tR{{\tilde R}}   \def\cL{{\cal L}}     \def\tcL{{\tilde\cL}}
\def\tom{{\tilde\omega}}

\def\nn{\nonumber}
\def\be{\begin{equation}}             \def\ee{\end{equation}}
\def\ba#1{\begin{array}{#1}}          \def\ea{\end{array}}
\def\bea{\begin{eqnarray} }           \def\eea{\end{eqnarray} }
\def\beann{\begin{eqnarray*} }        \def\eeann{\end{eqnarray*} }
\def\beal{\begin{eqalign}}            \def\eeal{\end{eqalign}}
\def\lab#1{\label{eq:#1}}             \def\eq#1{(\ref{eq:#1})}
\def\bsubeq{\begin{subequations}}     \def\esubeq{\end{subequations}}
\def\bitem{\begin{itemize}}           \def\eitem{\end{itemize}}

\title{``Exotic" black holes with torsion}

\author{M. Blagojevi\'c}
\email{mb@ipb.ac.rs}
\affiliation{University of Belgrade, Institute of Physics, P. O. Box 57,
11001 Belgrade, Serbia}

\author{B. Cvetkovi\'c}
\email{cbranislav@ipb.ac.rs}
\affiliation{University of Belgrade, Institute of Physics, P. O. Box 57,
11001 Belgrade, Serbia}

\author{M. Vasili\'c}
\email{mvasilic@ipb.ac.rs}
\affiliation{University of Belgrade, Institute of Physics, P. O. Box 57,
11001 Belgrade, Serbia}
\date{\today}

\begin{abstract}
In the context of three-dimensional gravity with torsion, the concepts of
standard and ``exotic" Ba\~nados--Teitelboim--Zanelli black holes are
generalized by going over to black holes with torsion. This approach
provides a unified insight into thermodynamics of black holes, with or
without torsion.
\end{abstract}

\maketitle

\section{Introduction}

Recently, Townsend and Zhang \cite{x1} examined thermodynamics of
``exotic" Ba\~nados--Teitelboim-Zanelli (BTZ) black holes---the solutions
of a class of 3D gravity models for which the metric \emph{coincides} with
the standard BTZ metric \cite{x2}, but the conserved charges, energy and
angular momentum, are, in a sense, \emph{reversed} (as explained in
chapter \ref{ch3}). Their analysis was focused on a simple model of this
type, described by the parity-odd gravitational Lagrangian that Witten
\cite{x3} named ``exotic". In this framework, the authors discussed basic
thermodynamic properties of the ``exotic" BTZ black holes (that is, the
standard BTZ black holes viewed as solutions of the ``exotic" model).

In Ref. \cite{x1}, general relativity with a cosmological constant (\grl)
and the ``exotic" gravity are treated as independent models, based on
\emph{Riemannian} geometry of spacetime. In  the present note, we show
that these two models can be naturally interpreted as different sectors of
a single model---the Mielke--Baekler (MB) model of \emph{3D gravity with
torsion} \cite{x4}. This approach offers a unified view at \grl\ and the
``exotic" gravity, revealing a new, ``interpolating" role of torsion with
respect to Riemannian theories of gravity. In this, more general setting,
standard BTZ black hole solutions can be generalized to BTZ-like
\emph{black holes with torsion} \cite{x5,x6,x7}, see also \cite{x8}. At
the same time, their thermodynamic properties \cite{x9,x10} allow us not
only to simplify the considerations presented in \cite{x1}, but also to
generalize them.

\section{3D gravity with torsion}

In the framework of Poincar\'e gauge theory \cite{x11,x12,x13}, where the
triad $e^i$ and the Lorentz connection $\om^i$ are basic dynamical
variables (1-forms), and their field strengths are the torsion
$T^i=de^i+\ve^{ijk}e_je_k$ and the curvature
$R^i=d\om^i+\frac{1}{2}\ve^{ijk}\om_j\om_k$ (2-forms), the MB model is
defined by the Lagrangian (3-form)
\be
L_\mb=2ae^iR_i-\frac{\L}{3}\ve_{ijk}e^ie^je^k
      +\a_3 L_\cs(\om)+\a_4 e^i T_i\, .                         \lab{1}
\ee
Here, $L_\cs(\om):=\om^id\om_i+\frac{1}{3}\ve_{ijk}\om^i\om^j\om^k$ is the
Chern--Simons Lagrangian for $\om^i$, the exterior product is omitted for
simplicity, and $(a,\L,\a_3,\a_4)$ are free parameters. In the
non-degenerate case $\a_3\a_4-a^2\ne 0$, the variation of $L_\mb$ with
respect to $e^i$ and $\om^i$ leads to the gravitational field equations in
vacuum:
\be
2T^i=p\ve^i{}_{jk}\,e^j\wedge e^k\, ,\qquad
  2R^i=q\ve^i{}_{jk}\,e^j\wedge e^k\, ,                         \lab{2}
\ee
where
\be
p=\frac{\a_3\L+\a_4 a}{\a_3\a_4-a^2}\, ,\qquad
  q=-\frac{(\a_4)^2+a\L}{\a_3\a_4-a^2}\, .                      \lab{3}
\ee
Using Eqs. \eq{2} and the formula $\om^i=\tom^i+K^i$, where $\tom^i$ is
the Riemannian (torsionless) connection, and $K^i$ is the contortion
1-form, defined implicitly by $T_i=\ve_{imn}K^m e^n$, one can show
\cite{x6,x14} that the Riemannian piece of the curvature, $\tR=R(\tom)$,
reads:
\be
2\tR^i=\Leff\,\ve^i{}_{jk}e^je^k\,,\qquad \Leff:=q-\frac{1}{4}p^2\,,\lab{4}
\ee
where $\Leff$ is the effective cosmological constant.

In the anti-de Sitter (AdS) sector with $\Leff=-1/\ell^2$, the MB model
admits a new type of black hole solutions, known as the BTZ-like
\emph{black holes with torsion} \cite{x5,x6,x7}. These solutions can be
determined in two steps. First, by combining the form the BTZ black hole
metric,
\bea
&&ds^2=N^2dt^2-N^{-2}dr^2-r^2(d\vphi+N_\vphi dt)^2\, ,          \nn\\
&&N^2=\left(-8Gm+\frac{r^2}{\ell^2}+\frac{16G^2j^2}{r^2}\right)\,,
  \quad N_\vphi=\frac{4Gj}{r^2}\, ,                             \nn
\eea
with the relation $ds^2=\eta_{ij}e^ie^j$, one concludes that the triad
field can be chosen in the simple, diagonal form:
\bsubeq\lab{5}
\be
e^0=Ndt,\quad e^1=N^{-1}dr,\quad e^2=r\left(d\vphi+N_\vphi dt\right)\,.
\ee
Then, the connection is determined by the first field equation in \eq{2}:
\be
\om^i=\tom^i+\frac{p}{2}e^i\, .
\ee
\esubeq
The pair $(e^i,\om^i)$ determined in this way represents the BTZ-like
black hole with torsion \cite{x5,x6,x7}. The thermodynamic aspects of the
new black holes are given as follows.

\emph{Energy and angular momentum} of the black hole with torsion, defined
as the on-shell values of the asymptotic generators for time translations
and spatial rotations, have the following form \cite{x5,x14}:
\bea
&&E=16\pi G\left[\left(a+\frac{\a_3p}{2}\right)m
                 -\frac{\a_3}{\ell^2}j\right]\, ,               \nn\\
&&J=16\pi G\left[\left(a+\frac{\a_3p}{2}\right)j
                 -\a_3 m\right]\, .                             \lab{6}
\eea
In contrast to \grl, where $E=m$ and $J=j$, the presence of the
Chern--Simons term ($\a_3\ne 0$) modifies $E$ and $J$ into linear
combinations of $m$ and $j$.

After choosing the AdS asymptotic conditions, the Poisson bracket algebra
of the \emph{asymptotic symmetry} is given by two independent Virasoro
algebras with different central charges \cite{x6,x14}:
\be
c^\mp=24\pi
  \left[\left(a+\frac{\a_3p}{2}\right)\ell \mp\a_3\right]\,.    \lab{7}
\ee

The partition function of the MB model, calculated in the semiclassical
approximation around the black hole with torsion, yields the following
expression for the \emph{black hole entropy} \cite{x9}:
\be
S=8\pi^2\left[\left(a+\frac{\a_3p}{2}\right)r_+
              -\a_3\frac{r_-}{\ell}\right]\, ,                  \lab{8}
\ee
where $r_\pm$ are the outer and inner horizons of the black hole, defined
as the zeros of $N^2$. The gravitational entropy \eq{8} coincides with the
corresponding statistical entropy \cite{x10}, obtained by combining
Cardy's formula with the central charges \eq{7}. The existence of torsion
is shown to be in complete agreement with the first law of black hole
thermodynamics.

\section{A special case: the results of Townsend and Zhang}\label{ch3}

After clarifying basic thermodynamic aspects of black holes with torsion,
the two types of black holes discussed in \cite{x1} can be given a unified
treatment by considering the related limiting cases of the MB model.

For $\a_3=\a_4=0$ and $16\pi Ga=1$, the MB model reduces to \grl, the
spacetime geometry is Riemannian ($p=0$), and formulas \eq{6}, \eq{7} and
\eq{8} produce the standard expressions for the conserved charges, central
charges and entropy:
\be
E=m,\quad J=j,\quad
c^\mp=\frac{3\ell}{2G},\quad S=\frac{2\pi r_+}{4G}.            \lab{9}
\ee

Similarly, for $a=\L=0$, the MB model reduces to Witten's ``exotic"
gravity with Riemannian geometry of spacetime. By choosing $16\pi
G\a_3=-\ell$, one arrives at the ``exotic" conserved charges, central
charges and entropy,
\be
E=\frac{j}{\ell}\,, \quad J=\ell m\,, \quad
  c^\mp=\pm\frac{3\ell}{2G}\,, \quad S=\frac{2\pi r_-}{4G}\,,   \lab{10}
\ee
which coincide with those in \cite{x1}. Since $\Leff=-1/\ell^2$ implies
$16\pi G\a_4=-1/\ell$, the corresponding ``exotic" Lagrangian is also the
same as in \cite{x1}.

These considerations, based on our earlier studies of black holes with
torsion, provide a simple way to understand somewhat enigmatic relation
between the standard and ``exotic" black hole thermodynamics.

\section{Generalization: Standard and ``exotic" black holes with torsion}

In the previous section, the concepts of standard and ``exotic" black
holes are used in the context of simple gravitational models with
Riemannian geometry of spacetime. Here, we wish to generalize these
concepts by going over to black holes with torsion.

The form of the general results \eq{6}, \eq{7} and \eq{8} suggests to
introduce \emph{standard} black holes with torsion by imposing the
following requirements:
\be
\a_3=0\, ,\qquad 16\pi G a=1\, .                                \lab{11}
\ee
In this case, the general formulas reduce to the standard form \eq{9}, and
the corresponding 2-parameter Lagrangian is given by:
\be
L_{\rm S}=\frac{1}{8\pi G}e^iR_i
      -\frac{\L}{3}\ve_{ijk}e^ie^je^k+\a_4 e^i T_i\, .          \lab{12}
\ee
The AdS condition,
\be
\Leff=\frac{3}{4}\left(\frac{\a_4}{a}\right)^2+\frac{\L}{a}
     =-\frac{1}{\ell^2}\, ,\nn
\ee
implies $\L<0$.

Similar considerations lead to the following definition of
\emph{``exotic"} black holes with torsion:
\be
a+\frac{\a_3 p}{2}=0\, ,\qquad 16\pi G\a_3=-\ell\, ,            \lab{13}
\ee
which implies that the conserved charges, central charges and entropy take
the ``exotic" form \eq{10}. The corresponding 2-parameter Lagrangian can be
written in the form
\bea
L_{\rm E}&=&\frac{1}{16\pi G}\left[2\b e^iR_i
  +\frac{\b(\b^2+3)}{3\ell^2}\ve_{ijk}e^ie^je^k\right.\nn\\
  &&\left.-\ell L_{CS}-\frac{\b^2+1}{\ell}e^i T_i\right]\, ,    \lab{14}
\eea
where $\b:=16\pi Ga$ and $\ell$ are free parameters.

In the limit $p = 0$, $L_{\rm S}$ and $L_{\rm E}$ describe torsionless
theories discussed by Townsend and Zhang [1]; thermodynamic aspects of the
corresponding black holes are given in Eqs. \eq{9} and \eq{10}. All the
other limits define the standard and ``exotic" gravities \emph{with
torsion}. In particular, for the choice $q=0$ (that is, by taking
$(\a_4)^2+\L/16\pi G=0$ in $L_{\rm S}$ and $\b=1$ in $L_{\rm E}$), the
geometry of these models becomes \emph{teleparallel} ($R^i=0$).

\acknowledgments

M.B. thanks F. W. Hehl for bringing the paper [1] to his attention. We
acknowledge the support from Grant No. 171031 of the Serbian Science
Foundation.


\end{document}